\documentclass[modern]{aastex62}
\usepackage{natbib}

\newcommand{\sourcefull}{GS 1826$-$238}
\newcommand{\source}{GS 1826}

\shorttitle{mHz Oscillations in \sourcefull}
\shortauthors{Strohmayer et al.}

\begin{document}

\title{NICER Discovers mHz Oscillations in the ``Clocked'' Burster GS
  1826-238}

\author{T. E. Strohmayer } 
\affil{Astrophysics Science Division and Joint Space-Science Institute,
  NASA's Goddard Space Flight Center, Greenbelt, MD 20771, USA}

\author{K. C. Gendreau}
\affil{X-ray Astrophysics Laboratory,
  Astrophysics Science Division, NASA's Goddard Space Flight Center,
  Greenbelt, MD 20771, USA} 

\author{D. Altamirano}
\affil{Physics \& Astronomy, University of Southampton, Southampton,
  Hampshire SO17 1BJ, UK}

\author{Z. Arzoumanian} 
\affil{X-ray Astrophysics Laboratory,
 Astrophysics Science Division, NASA's Goddard Space Flight Center,
  Greenbelt, MD 20771, USA} 


\author{P. M. Bult} 
\affil{Astrophysics Science Division, NASA's Goddard
  Space Flight Center, Greenbelt, MD 20771, USA}

\author{D. Chakrabarty} 
\affil{MIT Kavli Institute for Astrophysics
 and Space Research, Massachusetts Institute of Technology,
  Cambridge, MA 02139, USA}

\author{J. Chenevez} 
\affil{National Space Institute, Technical University of Denmark, 
Elektrovej 327-328, DK-2800, Lyngby, Denmark}


\author{S. Guillot} 
\affil{CNRS, IRAP, 9 avenue du Colonel Roche, BP
  44346, F-31028 Toulouse Cedex 4, France} 
\affil{Universit\'e de Toulouse, CNES, UPS-OMP, F-31028 Toulouse, France}

\author{T. Guver}
\affil{Istanbul University, Faculty of Science, Department of
  Astronomy and Space Sciences, Beyaz\i t, 34119, Istanbul,
  Turkey}
\affil{Istanbul University Observatory Research and Application
  Center, Beyaz\i t, 34119 Istanbul, Turkey}



\author{J. Homan} 
\affil{Eureka Scientific, Inc., 2452 Delmer Street,
  Oakland, CA 94602, USA}
\affil{SRON, Netherlands Institute for Space Research, 
Sorbonnelaan 2, 3584 CA Utrecht, The Netherlands}

\author{G. K. Jaisawal}
\affil{National Space Institute, Technical University of Denmark, 
Elektrovej 327-328, DK-2800, Lyngby, Denmark}

\author{L. Keek} 
\affil{Department of Astronomy, University of Maryland College Park, 
MD 20742, USA}


\author{S. Mahmoodifar} 
\affil{Astrophysics Science Division and Joint
  Space-Science Institute, NASA's Goddard Space Flight Center,
  Greenbelt, MD 20771, USA}

\author{J. M. Miller}
\affil{Department of Astronomy, University of Michigan, 1085 South University
Ave, Ann Arbor, MI 48109-1107, USA}


\author{F. Ozel} 
\affil{Astronomy Department, University of Arizona,
  933 N. Cherry Ave., Tucson, AZ 85721, USA}






\begin{abstract}

We report the discovery with the {\it Neutron Star Interior
  Composition Explorer} ({\it NICER}) of mHz X-ray brightness
oscillations from the ``clocked burster'' \sourcefull{}. {\it NICER}
observed the source in the periods 2017 June 20 - 29, July 11 - 13,
and September 9 - 15, for a total useful exposure of 34 ks. Two
consecutive dwells obtained on 2017 September 9 revealed highly
significant oscillations at a frequency of 8 mHz. The fractional,
sinusoidal modulation amplitude increases from $0.7 \%$ at 1 keV to
$\approx2\%$ at 6 keV. Similar oscillations were also detected at
lower significance in three additional dwells. The oscillation
frequency and amplitude are consistent with those of mHz QPOs reported
in other accreting neutron star systems.  A thermonuclear X-ray burst
was also observed on 2017 June 22. The burst properties and X-ray
colors are both consistent with \source{} being in a soft spectral
state during these observations, findings that are confirmed by
ongoing monitoring with {\it MAXI} and {\it SWIFT}-BAT.  Assuming that
the mHz oscillations are associated with black body emission from the
neutron star surface, modeling of the phase-resolved spectra shows
that the oscillation is consistent with being produced by modulation
of the temperature component of this emission. In this interpretation,
the black body normalization, proportional to the emitting surface
area, is consistent with being constant through the oscillation cycle.
We place the observations in the context of the current theory of
marginally stable burning and briefly discuss the potential for
constraining neutron star properties using mHz oscillations.


\end{abstract} 
\keywords{stars: neutron --- stars: oscillations --- stars: rotation ---
X-rays: binaries --- X-rays: individual (GS 1826$-$238) --- methods:
data analysis}

\section{Introduction}
\label{sec:introduction} 

Neutron stars in accreting X-ray binaries are known to exhibit
phenomena associated with unstable nuclear burning on their surfaces.
At present we know of 111 such
systems\footnote{https://personal.sron.nl/\textasciitilde
  jeanz/bursterlist.html} \citep{2017arXiv171206227G}.  These objects
produce hydrogen and helium-powered thermonuclear X-ray flashes, also
known as Type I X-ray bursts \citep{2006csxs.book..113S}, as well as
the longer and rarer ``superbursts,'' likely powered by carbon burning
\citep{2002ApJ...566.1045S, 2001ApJ...559L.127C}.  A much smaller
number of objects, five at present count, have shown low-frequency
($\approx 7-10$ mHz) oscillations very likely associated with the
transition from stable to unstable burning, known as marginally stable
nuclear burning \citep{2007ApJ...665.1311H, 2017arXiv171206227G}.

\cite{2001A&A...372..138R} reported the first detections of mHz
quasiperiodic oscillations (QPOs) in three accreting neutron star
systems, 4U 1608$-$52, 4U 1636$-$536, and Aql X-1. They argued that
the mHz oscillations were very likely associated with the neutron star
surface and not the accretion flow.  They also showed that they only
occur within a narrow range of X-ray luminosity.  Since then, mHz
oscillations have also been reported from 4U 1323$-$619
\citep{2012AAS...21924903S}, and the 11 Hz pulsar IGR J17480$-$2446 in
the globular cluster Terzan 5 \citep{2012ApJ...748...82L}. We note
that this object showed different behavior in its mHz oscillations
compared to the others. During its outburst in 2011, as the flux
increased, the recurrence time of bursts steadily decreased to about 3
minutes, finally appearing as oscillations in the light curve with
this period.  These oscillations also appeared at a much higher
inferred accretion rate than in the other mHz sources
\citep{2012ApJ...748...82L, 2017arXiv171206227G}.

\cite{2002ApJ...567L..67Y} found that the frequency of kilohertz QPOs
are anticorrelated with the brightness of the mHz oscillations in 4U
1608$-$52, and because this relationship is opposite to the positive
correlation of kHz QPO frequency with X-ray luminosity typically seen,
argued that the flux variation associated with the mHz QPO must arise
on the neutron star surface and not within the accretion
flow. \cite{2008ApJ...673L..35A} reported on extensive {\it Rossi
  X-ray Timing Explorer} ({\it RXTE}) observations of 4U 1636$-$536
that revealed a close connection between the mHz oscillations and the
occurrence of X-ray bursts. They found that, during intervals of mHz
oscillations, as the frequency drifted down and dropped below about
7.5 mHz, the oscillations faded and X-ray bursting resumed, and,
similarly to \cite{2001A&A...372..138R}, they found no mHz
oscillations immediately following the X-ray bursts.

More recently, \cite{2015MNRAS.454..541L} also studied episodes of mHz
QPOs in 4U 1636$-$536 using {\it XMM-Newton} and {\it RXTE} data.
Similarly to \cite{2008ApJ...673L..35A}, they found frequency drift of
the mHz oscillations, and suggested the drift time-scale may be set by
cooling of the deeper layers, as was previously argued by
\cite{2009A&A...502..871K} based on hydrodynamic calculations of
helium burning with rotational mixing.
\cite{2016MNRAS.463.2358L} found that the bursts that occurred
immediately after an episode of mHz oscillations preferentially showed
a rising light curve shape with so-called positive ``convexity.''
Such bursts show fast rise times, and this has been linked to a burst
ignition location at or near the star's rotational equator
\citep{2008MNRAS.383..387M, 2007ApJ...657L..29C}.  Thus,
\cite{2016MNRAS.463.2358L} argued that these bursts, and the burning
responsible for the mHz QPOs, may occur at the neutron star equator.
\cite{2016ApJ...831...34S} reported results from phase-resolved
spectroscopy of the mHz oscillations in 4U 1636$-$536 obtained with
{\it XMM-Newton}. They found an approximately constant color
temperature with pulse phase of $\approx 0.7$ keV that they associated
with the neutron star surface, and argued that the oscillations result
from a variation in the surface emitting area and not the temperature.
They derived a surface emitting area consistent with that expected for
a neutron star.

The physics relevant to marginally stable burning was recognized early
on by \cite{1983ApJ...264..282P}, and has been further elucidated by
\cite{2007ApJ...665.1311H}.  As the helium burning stability boundary
is approached, the temperature dependence of the nuclear heating rate
almost precisely balances that of the cooling rate. The result is a
slow, quasiperiodic mode of burning, with the oscillation period given
approximately by the geometric mean of the accretion and thermal
time-scales. For typical neutron star parameters and the relevant
accretion rates, this gives an oscillation period close to 2 min,
which is in good agreement with the range of oscillation frequencies
reported for the mHz oscillations.

Theory predicts that for unstable burning associated with helium
ignition, the boundary between stable and unstable burning should
occur at local accretion rates close to the Eddington rate
\citep{1998ASIC..515..419B}.  The fact that observational indications
based on estimates of the X-ray luminosity suggest the onset of
stability at accretion rates closer to $\approx1/10$ of the Eddington
rate remains a major puzzle \citep{1988MNRAS.233..437V,
  2003A&A...405.1033C}.  It has been suggested that confinement of the
accreted fuel, and/or rotational mixing of the fuel to greater column
depths could alleviate this apparent discrepancy
\citep{2009A&A...502..871K}. The fact that the mHz oscillations
associated with marginal stability occur at a well defined local
accretion rate makes them, in principle, a particularly important
probe of the accretion rate. \cite{2014ApJ...787..101K} found that the
range of accretion rates in which marginally stable oscillations can
occur was also sensitive to the nuclear reaction rates.  Moreover,
\cite{2007ApJ...665.1311H} showed that the oscillation frequency
associated with marginal stability is sensitive to the hydrogen mass
fraction in the accreted fuel, as well as the surface gravity.  Thus,
a more detailed understanding of these oscillations could lead to new
probes of these quantities.

The accreting neutron star binary \sourcefull{} (hereafter, \source{})
is in many ways a prototype, having earned the moniker ``clocked
burster,'' for producing bursts with extreme regularity
\citep{1999ApJ...514L..27U}. The recurrence times of these regular
bursts were shown to match well the predictions of theory for mixed
H/He bursts \citep{2004ApJ...601..466G}, and the light curve shapes
were accurately modeled using theoretical calculations of rapid-proton
(rp)-process burning of approximately solar composition fuel
\citep{2007ApJ...671L.141H}, hence, it has also been referred to as
the ``text-book'' burster.  Indeed, the bursts were so uniform in
their properties that \cite{2012ApJ...749...69Z} used them as
``standard candles'' to place constraints on the mass and radius of
the neutron star.

The ``clocked'' bursts all occurred, however, while \source{} was in a
``hard'' spectral state.  This is typically diagnosed with the use of
X-ray colors \citep{2006csxs.book...39V}.  Based on this, \source{} is
classified as an ``atoll'' source, and the hard state is also referred
to as the ``island'' state.  Interestingly, \cite{2016ApJ...818..135C}
recently reported on observations of \source{} in 2014 June with {\it
  Swift} and {\it NuSTAR} during a soft spectral state, including
the detection of several X-ray bursts.  They found that in this
spectral state the X-ray burst recurrence times were no longer
regular, and the bursts themselves also differed from the hard state
bursts, being generally shorter in duration.  These findings are
indicative of a lower hydrogen fraction in the fuel, and suggest an
additional source of stable burning occurs at the higher accretion
rates in the soft state.  They also detected the first photospheric
radius expansion (PRE) burst from \source{}, and used it to estimate
the distance as $5.7 \pm 0.2$ kpc (assuming isotropic emission). Since
then, the source has mostly remained in a soft state.

In this paper we report the discovery with {\it NICER} of mHz
oscillations from \source{} that are very likely associated with
marginally stable burning.  The plan of the paper is as follows. We
begin with a description of the observations and present the detection
of mHz oscillations, and an X-ray burst. Next, we explore the source
accretion state by studying the X-ray colors, showing that the source
was in a soft, ``banana'' state in the color - color diagram
\citep{1989A&A...225...79H}.  Next we study the average mHz pulse
profile, and present the results of phase-resolved spectroscopy.  We
conclude with a brief summary and discussion of the implications of
our findings for models of marginally stable burning and for
constraining the neutron star properties of \source{}.

\section{NICER Observations}

\label{sec:observations} 
{\it NICER} was installed on the International Space Station ({\it
  ISS}) in 2017 June, and provides low background, high throughput
($\approx 1900$ cm$^2$ at 1.5 keV), fast timing observations across
the $0.2 - 12$ keV X-ray band \citep{2012SPIE.8443E..13G}. {\it NICER}
achieves an absolute timing precision of $\approx 100$ ns with the aid
of a GPS receiver. {\it NICER} observed \source{} as part of a science
team program with a major goal to detect and study X-ray
bursts. Observations were obtained in the periods 2017 June 20 - 29,
July 11 - 13 and September 9 - 15. Relevant ``observation IDs''
(obsids) include $00503101nn$ and $10503101mm$, where $nn$ runs from
01 to 14 and $mm$ from 01 to 16.  We used {HEASOFT} version 6.23 and
NICERDAS version 2018-03-01\_V003 to process the data, and we applied
the standard filtering criteria, which includes time intervals with a
pointing offset $<0.015^{\circ}$
\footnote{We note that obsids $0050310101-05$, inclusive, were
  obtained with an offset pointing of 0.03$^{\circ}$, which nominally
  violates the standard filtering criterion. However, this offset
  simply reduces the count rate modestly but does not otherwise effect
  our results, so we included these data in our analysis.}, bright
Earth limb angles $> 40^{\circ}$, dark Earth limb angles
$>30^{\circ}$, and outside the South Atlantic Anomaly (SAA).  This
resulted in a total good exposure of 18, 6 and 10 ks for the June,
July, and September epochs, respectively.

\subsection{mHz QPOs and an X-ray Burst}

We found that a few data intervals showed enhanced backgrounds at
energies above $\approx 8$ keV, so to mitigate this we restricted the
timing analysis to events with energies less than 7.5 keV.  We note
that, because the observed count rate is decreasing with energy, only
$\approx 0.5 \%$ of all observed events have energies greater than
this limit. We computed light curves for each epoch using events in
the range $0.4 - 7.5$ keV.  Figure 1 shows the resulting light curve,
where the time axis is broken up into three panels, one each for the
June, July, and September exposures, respectively.  We identified a
thermonuclear X-ray burst that occurred on 2017 June 22  at 22:31:42
UTC (labelled in Figure 1).  The burst shows a hardness ratio
evolution consistent with photospheric radius expansion (PRE), and
also shows a shorter-duration light curve that differs from those of
the hard state, ``clocked'' bursts \citep{2004ApJ...601..466G,
  2007ApJ...671L.141H}, but is similar to bursts recently observed
from \source{} in a high (soft) spectral state
\citep{2016ApJ...818..135C}.  We note that the publicly available data
from the {\it Monitor of All-sky X-ray Image} ({\it MAXI}) and the
{\it Swift} {\it Burst Alert Telescope} ({\it BAT}), are consistent
with \source{} being predominantly in a soft spectral state from about
2016 December to the present.  We will present a detailed spectral
study of this burst in a subsequent paper.

Visual inspection of the light curves from the first two dwells of the
September data epoch (these dwells are the first in the right-most
panel of Figure 1, and are also labelled ``mHz QPO'') suggested the
presence of apparent pulsations with a few minute period.  The light
curves of these dwells in the $0.4 - 7.5$ keV band, binned at 10 s,
are shown in Figure 2.  Pulses are rather clearly evident ``by eye.''
To explore this apparent variability more quantitatively we computed
power spectra using continuous intervals of 2200 s for each of these
dwells.  We again used events in the $0.4 - 7.5$ keV band, and sampled
the light curves at 8192 Hz.  The resulting power spectra, normalized
as in \cite{1983ApJ...266..160L}, are shown in Figure 3, where the
black and red curves correspond to the first and second dwells,
respectively.  Both spectra show strong, narrow peaks near 8 mHz, a
frequency consistent with the few minute separation of pulses evident
in Figure 2. These peaks are highly significant, for pure Poisson
noise the chance probability to exceed a power value of 200 is $3.7
\times 10^{-44}$.  From the power spectra shown in Figure 3 we
estimate fractional amplitudes (rms) in the $0.4 - 7.5$ keV band for
the first and second dwells of $0.79\%$ and $0.95\%$, respectively.
These values for the frequency and amplitude are largely consistent
with those previously reported for mHz oscillations in other sources
\citep{2001A&A...372..138R}.

The mHz signal power from the first dwell is confined to a single
Fourier bin, being effectively coherent over the length of the
interval.  The Fourier frequency of this bin is 8.18 mHz.  In the
second dwell the peak power shifts down one Fourier bin, and
significant power is spread over a few bins.  This suggests a downward
drift in the oscillation frequency from the first to second dwell. To
quantify this drift we oversampled the power spectrum for each dwell
by padding the light curve to 8000 s using the mean count rate
determined from the good exposure (2200 s). We then found the mHz
frequency associated with the peak power in each spectrum.  From the
frequency separation, $\Delta\nu$, of the peak power values we
estimate a frequency drift rate of $\Delta\nu / \Delta t \approx -0.3$
mHz hr$^{-1}$.  Previous studies of mHz oscillations indicate that the
frequency can both increase and decrease with time, and that the
frequency drift rate can vary \citep{2014MNRAS.445.3659L,
  2015MNRAS.454..541L, 2008ApJ...673L..35A}. The best-studied object
in this regard is 4U 1636$-$536.  The rate of decrease we estimate
here for \source{} does not appear to be inconsistent with previous
reports \citep{2008ApJ...673L..35A}, that is, one can find time
intervals reported in the literature for 4U 1636$-$536 which have
approximately similar drift rates (see, for example, Figure 4 in Lyu
et al 2015).

After identifying the strong mHz pulsations described above, we also
searched for similar signals in the remaining data. We note that many
of the other individual dwells have exposures that are shorter than
those of the 2017 September dwells with strong mHz oscillations. So,
not all observations are equally sensitive to such oscillations.
Nevertheless, we searched for additional signals by computing power
spectra of all dwells with exposures equal to or longer than 750 s
($0.4 - 7.5$ keV band).  We detected mHz oscillation signals near 8
mHz in three other dwells, while for the remainder no signal was
detected, and we obtain upper limits ($90\%$ confidence) to an
oscillation in the 7 - 9 mHz frequency band that range from
approximately $0.3 - 0.5\%$ (rms), with the precise value depending on
the count rate and exposure of the particular dwell.  Figure 4 shows
power spectra from these three additional dwells with mHz
oscillations.  The spectra are labelled with the start date (MJD) of
the dwell, and the number in parenthesis indicates the corresponding
day in Figure 1. The successive spectra are displaced vertically by
100 for clarity. From figure bottom to top we estimate fractional
amplitudes (rms) of $0.60$, $0.85$ and $0.65 \%$. We also searched for
mHz oscillations in the immediate aftermath of the X-ray burst
detected on 2017 June 22, but no mHz signal was detected in a power
spectrum computed from a 600 s interval ($0.4 - 7.5$ keV) following
the burst. The upper limit on the amplitude is $0.5 \%$
(rms). Finally, we did not detect mHz oscillations in any of the
dwells preceding the X-ray burst.

\subsection{Spectral State}


To further explore the accretion state context of the mHz oscillations
we calculated color - color and hardness - intensity diagrams for all
the data.  We computed both a ``soft'' and ``hard'' color. For the
former we determine the ratio of count rates in the bands $1.8 - 3.5$
and $0.5 - 1.8$ keV, while for the latter we use $5.2 - 6.8$ and $3.5
- 5.2$ keV.  For the intensity we compute the count rate from $0.5 -
6.8$ keV.  We computed colors and intensities using 32 s intervals for
all the data. Figures 5 and 6 show the resulting color - color, and
hardness - intensity diagrams, respectively.  In both figures the blue
(triangle) symbols represent data obtained from the two September
epoch intervals when strong mHz oscillations were present, and the red
(diamond) symbols were obtained from intervals immediately following
the X-ray burst. A characteristic error bar is also shown in each plot
to indicate the statistical precision.

To further elucidate the relationship between the spectral colors and
mHz oscillations we computed colors for all dwells in which
oscillations were detected, as well as some of the longer dwells with
no detections, and only upper limits. The colors measured for all of
these dwells are shown in Figure 7, where the red diamond symbols
represent those with upper limits, the blue square symbols those with
mHz oscillation detections, and the size of the blue symbols is
proportional to the oscillation amplitude (rms).  The X-ray colors are
defined in the same way as in Figure 5. These results suggest that the
presence and amplitude of mHz oscillations is related to position in
the color - color diagram, and therefore the mass accretion rate, as
dwells with the highest amplitudes seem to occur preferentially at the
higher values of soft and hard color.  Such behavior is theoretically
predicted for mHz oscillations associated with marginally stable
burning \citep{2007ApJ...665.1311H}, as they are only evident close to
the mass accretion rate at which burning stabilizes.  It may be that
we are seeing such a transition in Figure 7, since mHz oscillations
are not detected at the lowest inferred accretion rates (the lower
left portion of the color - color diagram). However, it is also
possible that this behavior is similar to that seen in 4U 1636$-$536,
where mHz oscillations can ``come and go'' while on the ``banana''
branch \citep{2008ApJ...673L..35A}. The present data are too sparse to
be conclusive, and more observations will be needed to distinguish
between these two possibilities.


\section{Properties of mHz Oscillations}

Since the power spectra (Figure 3) from the two 2017 September dwells
with strong mHz oscillations are consistent with rather coherent
oscillations during these intervals, we computed phases for all events
in each of these dwells separately using the frequencies corresponding
to the peak Fourier power obtained from each dwell.  To combine the
two dwells we first found a constant phase offset for the 2nd dwell
that maximized the $Z_1^2$ signal power of the sum of both dwells
\citep{1983A&A...128..245B}.  The resulting phase folded pulse profile
in the $0.4 - 7.5$ keV band is shown in Figure 8.  We fit a model of
the form $A + B\sin(\phi -\phi_0)$ and find an amplitude $B/A = 1.12
\pm 0.05 \%$. The fit is good with a minimum $\chi^2 = 11.0$ for 13
degrees of freedom, and the best fitting model is also plotted in
Figure 8 (red curve).  To explore the energy dependence we also
computed profiles for different energy bands and then fit them with
the same model as above.  Figure 9 shows the derived amplitudes,
$B/A$, and phases, $\phi_0$, versus energy.  The amplitude shows a
modest increase from about $0.7 \%$ at 1 keV to $\approx 2 \%$ at 7
keV.  There is no strong evidence for a variation in $\phi_0$ with
energy.

To explore the pulse-phase spectrum we extracted two spectra, from
regions around the pulse minimum and maximum. The phase ranges used
for these extractions are indicated by the dashed and dash-dotted
vertical lines in Figure 8, respectively.  In each case we used a
range of pulse phase equal to 0.25 cycles, so that each spectrum has
an exposure of $0.25\times(2(2200)) = 1100$ s.  We used XSPEC version
12.9.1 with version 1.02 of the {\it NICER} instrument response, and
we fit the spectra in the range $0.6 - 9$ keV. We use a model
comprised of thermal comptonization ({\it comptt} in XSPEC) for the
persistent accretion-driven emission and a black body ({\it bbodyrad}
in XSPEC) to capture the neutron star thermal emission
\citep{2016ApJ...818..135C, 2008ApJ...681..506T}. To model Galactic
absorption we employ the {\it tbabs} model with abundances from
\cite{2000ApJ...542..914W}, and we fix (freeze) the value of $n_{H} =
0.4 \times 10^{22}$ cm$^{-2}$ based on prior measurements, since the
source is in the same (soft) state \citep{2016ApJ...818..135C,
  2010A&A...521A..79P}. We employ a background spectrum extracted from
{\it NICER} observations of {\it RXTE} background field number 5
\citep{2006ApJS..163..401J}, but we note that the spectrum is strongly
source dominated across the fitted band.  Finally, we add a $1\%$
systematic error to accommodate current uncertainties in the {\it
  NICER} response.

As with many X-ray spectral modeling efforts, there is often no
single, unique approach. The same is true in this case. In order to
model either individual phase-resolved spectrum we find that both of
the spectral components discussed above are required. However, the
data do not have sufficient statistical quality to simultaneously fit
for all parameters and determine unambiguously whether either of the
components alone is associated with the mHz modulation. That is, one
can either fix the persistent component ({\it comptt}) to be the same
between the phase maximum and minimum spectra, or one can fix the
black body ({\it bbodyrad}) component. Given the statistical quality
of the present spectra one can achieve a similarly good fit using
either approach. Below, we explore the approach that the persistent
emission is not strongly affected by the mHz oscillation, and assume
that the difference between the two spectra is accounted for by the
black body component.  This appears reasonable given the modest $1 -
2\%$ amplitude of the modulation, and the theoretical suggestion that
such oscillations arise from the neutron star surface. However, we
emphasize that this approach places constraints on the black body
component {\it under the assumption that it is the varying component},
but it does not demonstrate unambiguously that it is the varying
component.

Thus, we carry out a joint fit to both the minimum and maximum phase
spectra, keeping the {\it comptt} parameters and the column depth,
$n_H$, tied together, but allowing the black body parameters to vary
independently between the two spectra.

\begin{table*}
\renewcommand{\arraystretch}{1.15}
\caption{Phase-resolved spectral fits for \source{} }
\scalebox{0.95}{
\begin{tabular}{cccc}
\tableline\tableline
Component & Parameter & Spectrum & Spectrum \\
          &           & minimum & maximum \\
\tableline
{\bf tbabs} & $n_{H}$ ($10^{22}$ cm$^{-2}$) & $0.4$ & tied  \\
{\bf compTT} & $kT_0$ (keV) & $0.095 \pm 0.014$ & tied \\
             & $kT_e$ (keV) & $2.03 \pm 0.04$ & tied \\
             & $\tau$       &  $8.4 \pm 0.1$   &  tied \\
             & {\it norm}   &  $1.88 \pm 0.20$   &  tied \\
{\bf bbodyrad} & $kT$ (keV)  &  $0.684 \pm 0.006$   &  $0.701 \pm 0.005$ \\
             & $K$ ({\it norm}) & $603.7 \pm 14.3$   & $599.0 \pm 13.1$ \\
\tableline
\end{tabular}}
\tablecomments{Parameter uncertainties are given at $1\sigma$
  confidence, and those with no quoted uncertainty were held fixed at
  the indicated value. }
\end{table*}

We find this model provides a reasonable description of the observed
spectra, with a reduced $\chi^2_r \approx 1.01$ ($\chi^2 = 1698$ with
1676 degrees of freedom). The largest remaining residuals are centered
near 2.3 keV, and this feature is associated with a known instrumental
edge (see \citet{2018ApJ...858L...5L} for further discussion of the
{\it NICER} spectral response). Our spectral results are summarized in
Table 1. Figure 10 shows the unfolded, photon spectra, including the
individual model components (top), and the fit residuals (bottom).
The mean (unabsorbed) flux in the $0.6$ - $9$ keV band is $6.90 \pm
0.01 \times 10^{-9}$ erg cm$^{-2}$ s$^{-1}$.  For the comptonization
component parameters we find results that are qualitatively similar to
\cite{2016ApJ...818..135C}, but differ somewhat in the details. For
the black body components we find that the fitted normalizations are
consistent with remaining constant from pulse minimum to maximum, with
a mean value of $K_{avg} = 601.3$.  This normalization corresponds to
an area, assuming a spherical emitting source and isotropic emission,
with radius $R = (K_{avg} d_{10}^{2} )^{1/2} = 13.98$ km, where
$d_{10}$ is the distance in units of 10 kpc, and we have assumed $d =
5.7$ kpc \citep{2016ApJ...818..135C}. We note that this estimate
ignores any atmospheric color corrections. Applying such a correction
would increase the inferred radius by a factor of $f_c^2 = (T_c /
T_{eff})^2$, where $T_c$ and $T_{eff}$ are the color (measured) and
effective temperatures, respectively, and $f_c$ is the so-called
hardening factor. Based on model atmosphere calculations, $f_c$ is
likely to fall within the range from $\approx 1.2 - 1.5$
\citep{2016ApJ...832..102M, 2011A&A...527A.139S}, resulting in an
increase in inferred radius by a factor of $\approx 1.4 - 2.2$.
However, given the other assumptions and uncertainties, the value is
at least approximately consistent with expectations for surface
emission from a neutron star.

For the spectra at maximum and minimum oscillation phase we find
temperature values of $kT = 0.701 \pm 0.005$ keV, and $kT = 0.684 \pm
0.006$ (both $1\sigma$ confidence), respectively.  This suggests that
the flux modulation associated with the mHz oscillation appears to be
consistent with a change in the surface temperature, $kT$, of the
neutron star thermal emission.  However, we note that these values
differ only at the $\approx 3\sigma$ level, and more data are needed
to demonstrate this conclusively.

\section{Discussion}

We report the detection with {\it NICER} of mHz oscillations in
\source{} for the first time.  The oscillation properties are
generally consistent with those of the mHz QPOs observed in 4U
1608$-$52, 4U 1636$-$536, Aql X-1, and 4U 1323$-$619.  The {\it NICER}
data were obtained while \source{} was in a soft spectral state, as
indicated by both {\it MAXI} \citep{2011PASJ...63S.635S} and {\it
  Swift/BAT} \citep{2013ApJS..209...14K} long-term monitoring from
about 2015 December to the present. Indeed, the current long-term
trend appears to be a slow but steady increase in the 2 - 10 keV flux
(based on the {\it MAXI} data), and very weak hard X-ray flux (from
{\it Swift}-BAT).  Our spectral color analysis appears consistent with
this conclusion as well.  The observed behavior (Figures 5, 6 and 7)
appears consistent with \source{} tracing out a portion of the
so-called ``banana'' branch in the color - intensity diagram during
the {\it NICER} observations.  The mass accretion rate is generally
inferred to increase from left to right along the arc of the
``banana.''  To further test this we extracted spectra from several
intervals which populate the lower left portion of the color - color
diagram (see Figure 7), in order to estimate their fluxes and compare
with the mean flux measured during the September epoch intervals with
strong mHz oscillations (see \S 2.2 above). We find these spectra can
be fit with the same model described above, and we find the unabsorbed
flux ($0.6$ - $9$ keV) in the range from $5.0 - 5.2 \times 10^{-9}$
erg cm$^{-2}$ s$^{-1}$, supporting the conclusion that the X-ray flux,
and therefore mass accretion rate, increases from lower left to upper
right in the color - color diagram (Figure 7).

Assuming the X-ray luminosity can be written as,
\begin{equation}
 L_x = \frac{GM \dot M}{(1+z)R} = 4\pi d^2 f_x \; ,
\end{equation}
  where $M$, $R$, $(1+z) = (1 - 2GM/c^2 R)^{-1/2}$, and $f_x$ are the
  neutron star mass, radius, surface redshift, and X-ray flux
  respectively, we estimate an average mass accretion rate, $\dot M$,
  for the September dwells with mHz oscillations of $\dot M = 0.18
  \dot M_{Edd}$, where we have assumed $M = 1.4 M_{\odot}$, $R = 10$
  km, $d = 5.7$ kpc, and $\dot M_{Edd} = 1.7 \times 10^{-8} M_{\odot}$
  yr$^{-1}$. This estimate also assumes that the accretion luminosity
  is radiated isotropically.

Previous studies have shown that mHz oscillations can be
  associated with bursting activity.  For example,
  \cite{2008ApJ...673L..35A} found episodes of mHz oscillations in 4U
  1636$-$536 with decreasing frequency, and when the frequency dropped
  below about 7.5 mHz the oscillations faded and bursting resumed;
  however, such behavior was not associated with {\it all} X-ray
  bursts (see their Figure 3). Indeed, the data for 4U 1636$-$536
  suggest that intervals of mHz oscillations can come and go without
  an obvious trigger, as long as the source is in the appropriate
  accretion state, indicated by its position on the banana branch or
  in the transition to the banana branch (see Altamirano et al. 2008,
  Figure 1).  For \source{} we find no evidence for mHz oscillations
  after the X-ray burst on 22 June, which at face value is consistent
  with the behavior seen in 4U 1636$-$536. However, we emphasize that
  the present data are very sparse around the time of the X-ray
  burst. The only thing we can say with certainty is that we did not
  detect mHz oscillations in any of the {\it NICER} dwells preceeding
  the X-ray burst or immediately after it.

Our results suggest the following long term behavior in
\source{}. Prior to 2014 the source was almost exclusively in the hard
state, with typical accretion rates lower than indicated by the
present {\it NICER} observations.  For example, comparing long term
variations in the persistent X-ray flux--a reasonable proxy for mass
accretion rate--we find from \cite{2004ApJ...601..466G} that the $2.5
- 25$ keV persistent (absorbed) flux from {\it RXTE} observations was
$2.2 \times 10^{-9}$ erg cm$^{-2}$ s$^{-1}$ on July 29, 2002.  Based
on the {\it NICER} spectral modeling for the 2017 September epochs
described above we estimate a flux in the same band of $4.5 \times
10^{-9}$ erg cm$^{-2}$ s$^{-1}$, a factor of two or so higher.  In
making this estimate we note that we have extrapolated the best-fit
{\it NICER} model (Table 1) to 25 keV, that is, outside of {\it
  NICER}'s nominal bandpass.  However, the (absorbed) flux measured by
{\it NICER} in the $2.5 - 9$ keV band is $3.6 \times 10^{-9}$ erg
cm$^{-2}$ s$^{-1}$, which is already significantly larger than the
$2.5 - 25$ keV {\it RXTE} flux. Since the $2.5 - 9$ keV {\it NICER}
flux measurement is a firm lower bound to the $2.5 - 25$ keV flux,
this strongly suggests that the flux measured by {\it NICER} in 2017
September {\it in the same band} is significantly higher than the July
2002 flux measured with {\it RXTE}. This provides further support to
the notion that the apparent long-term increase in X-ray flux likely
represents an increasing mass accretion rate.

While in the hard state the source was extensively observed with {\it
  RXTE}, but no mHz QPOs were ever observed. As the source continued
to brighten and soften, the accretion rate became large enough to
enter the accretion regime where mHz oscillations are observed in
other LMXBs (see, for example, Altamirano et al. 2008, Figure 1). This
overall behavior appears consistent with the observations of the mHz
QPOs in 4U 1636$-$536, 4U 1608$-$52 and Aql X-1, and suggests we are
seeing the same phenomenon in \source{}.  More observations of
\source{} in its soft, higher accretion rate state will be needed to
determine if it shows the full range of mHz oscillation phenomenology
evident in other sources, including a closer connection with X-ray
bursts.

There have been some reports of QPOs not linked with marginally stable
burning, but with similar frequencies. In some of those cases the
accretor is known or strongly suspected to be a black hole and not a
neutron star. Examples are the QPOs identified in the black hole
candidates H1743$-$322 \citep{2012ApJ...754L..23A}, and LMC X-1
\citep{2014MNRAS.445.4259A}.  In the case of H1743$-$322, the QPOs are
seen in a hard spectral state, whereas the mHz QPOs reported here
occur in the soft state of \source{}.  For LMC X-1 the observed QPO
frequencies are higher, at $\approx 27$ mHz, than any previously
reported mHz oscillation associated with marginally stable burning.
Based on this it appears unlikey to us that the mHz oscillations
observed in \source{} are related to these QPOs.

Several accreting millisecond X-ray pulsars have also occasionally
shown episodes of QPOs with mHz frequencies during some outbursts, an
example being the 8 mHz ``flaring'' observed in IGR J00291+5934
\citep{2017MNRAS.466.3450F}.  In those QPOs, however, the modulation
amplitude was $> 10\%$ (rms) for photon energies less than 2 keV,
which is substantially larger than is found for the mHz oscillations
linked to marginally stable burning. Morever, the energy spectrum
softened at the peak of these flares, which is opposite to the
behavior indicated in \source{}.  This argues rather strongly that the
mHz oscillations reported here from \source{} are not related to these
mHz ``flaring'' QPOs.

As discussed in \S 2.1 the oscillations reported here from \source{}
have a rather high quality factor, that is, they appear to be rather
coherent over the intervals they are observed. We emphasize, however,
that these intervals ($\approx 2$ ks) are relatively short compared to
the hours-long time-scales over which mHz QPO frequencies have been
observed to drift in several sources \citep{2015MNRAS.454..541L,
  2008ApJ...673L..35A}.  Additionally, current modeling of marginally
stable burning can produce quite regular trains of mHz pulses, with
the limiting factor here likely being temporal variations of the mass
accretion rate \citep{2007ApJ...665.1311H, 2014ApJ...787..101K}.
Based on these arguments we think that the relatively high coherence
seen in the mHz oscillations reported here is not inconsistent with
their connection to marginally stable burning.

Figure 9 shows the energy dependence of the fractional (sinusoidal)
modulation amplitude of the mHz oscillations in \source{}.  From about
$1 - 3$ keV an increasing trend is rather clear, with an apparent
flattening above 3 keV.  This appears to be at least qualitatively
similar to the behavior seen in IGR J17480$-$2446--which shows an
increasing trend in rms amplitude from about $2.5 - 10$ keV based on
{\it RXTE} data \citep{2012ApJ...748...82L}--but opposite to that
reported for 4U 1608$-$52 and 4U 1636$-$536 by
\cite{2001A&A...372..138R}, also using {\it RXTE}.  However, some
caution is warranted here due to the different bandpasses of {\it
  NICER} and {\it RXTE}, as well as the extent to which uncertainties
in backgrounds could influence derived amplitudes.  A closer
comparison of the energy dependence of the mHz oscillations reported
here with {\it NICER}, and those detected with {\it XMM-Newton} should
prove useful, as these instruments have more closely matched
bandpasses.

We extracted spectra as a function of mHz pulsation phase, and find
that they are consistent with the modulation being produced by a
variation in the temperature of a thermal (black body) component at an
effectively constant emitting area, though we emphasize again that
this is not a unique spectral interpretation (see discussion in \S 3
above). Assuming a spherical emitting source and isotropic emission,
the fitted black body normalization implies a radius of $R \approx 14$
km.  Given uncertainties in the source distance, the exact nature of
the continuum describing the persistent emission, the extent to which
disk reprocessing is present, as well as our neglect of color
corrections, this value should only be taken as roughly indicative of
the surface area associated with the mHz oscillations, and definitely
not a precise measurement.  The evidence that the oscillating flux can
be associated with a thermal (black body) component with an area
approximately consistent with that for a neutron star provides support
for the interpretation of the oscillations as due to marginally stable
nuclear burning; however, more data will be needed to establish this
definitively.

We note that \cite{2016ApJ...831...34S} reached different conclusions
from a study of the phase-resolved spectra of mHz oscillations from 4U
1636$-$536.  They argued that the flux modulation resulted from a
variation in the emitting area at approximately constant
temperature. While this is at odds with current theory, which finds
that the temperature oscillates \citep{2007ApJ...665.1311H}, we
emphasize that current theoretical calculations of marginally stable
burning are all one dimensional, so they essentially do not explore
the possibility of variations in the burning area.  

Here we point out that there are several physical effects that suggest
that lateral variations in the burning should be included in future
modeling.  For example, \cite{2007ApJ...665.1311H} showed that the
oscillations associated with marginal stability occur in a very narrow
range of mass accretion rates, $\dot m$, for a given surface
gravity. They found the oscillations were present in a range of only
$1\%$ of the critical accretion rate.  Note that it is the {\it local}
accretion rate that is relevant for the nuclear burning. For a fast
spinning neutron star the variations in effective surface gravity from
equator to pole can result in a latitudinal change in the local
accretion rate.  \cite{2007ApJ...657L..29C} used this effect to
explore the extent to which ignition of thermonuclear bursts could
preferentially occur at different latitudes. However, this process
would appear to be relevant for marginally stable burning as well, as
a greater than $1\%$ variation in the local accretion rate with
latitude could influence the range of stellar latitudes for which
marginally stable oscillations could occur. Indeed, this appears to be
a possible mechanism for limiting the marginally stable burning to a
belt in latitude \citep{2016MNRAS.463.2358L}. As a rough guide, a star
rotating at 582 Hz (close to the inferred spin frequency for 4U
1636$-$536), and with a mass of $1.4 M_{\odot}$ and equatorial radius
of 10 km, would have an $\approx 11\%$ increase in effective surface
gravity from equator to pole, using the results of
\cite{2014ApJ...791...78A}. This is certainly large enough to have a
significant effect on the local accretion rate at the level required
to influence marginally stable burning.  Interestingly, for a spin
rate of 11 Hz, appropriate for IGR J17480$-$2446 in Terzan 5, the
variations in surface gravity are $< 0.004 \%$ equator to pole, and
thus latitudinal variations would likely be negligible for such a slow
spinner.  This could perhaps account for some of the observed
differences in the mHz oscillations of IGR J17480$-$2446 and fast
spinners like 4U 1608$-$52, 4U 1636$-$536, and Aql X-1.  As pointed
out by \cite{2007ApJ...665.1311H}, the critical accretion rate for
marginally stable oscillations is itself a function of surface
gravity, so the occurrence of oscillations, and their properties, will
in principle depend on both effects; however, this discussion
highlights that such latitudinal dependencies of the burning are
likely, at least for fast spinners, and should be considered in future
modeling of marginally stable oscillations.

\cite{2007ApJ...665.1311H} argued that the oscillation frequency
associated with marginally stable burning should be sensitive to the
neutron star surface gravity, which is proportional to $M/R^2$, and
the mass fraction of hydrogen in the accreted matter (see their Figure
9).  While their results on this score were based on one-zone models,
the good overall agreement between predictions of their one-zone
models and multi-zone hydrodynamic calculations with the KEPLER code
\citep{2007ApJ...665.1311H}, suggests that these dependencies are
robust.  This is particularly promising for \source{} because of the
extensive theoretical modeling already done on the ``clocked'' bursts
\citep{2007ApJ...671L.141H, 2018ApJ...860..147M}, which suggests a
near solar composition for the accreted fuel. Interestingly, the
bursts observed so far at the higher accretion rates associated with
the soft spectral state have shorter durations, and several have
reached higher peak fluxes, as is evidenced by the occurrence of PRE.
These are both good indications of depletion of hydrogen in the fuel
compared to the hard-state, ``clocked'' bursts.  This suggests that
additional theoretical modeling of the soft state bursts, coupled with
further comparisons to the ``clocked'' bursts, could place tighter
constraints on the hydrogen fraction present in the fuel during
episodes of mHz oscillations.  It might then be possible to provide
more robust limits on the neutron star surface gravity from
measurements of the mHz oscillations.

\acknowledgments

This work was supported by NASA through the {\it NICER} mission and
the Astrophysics Explorers Program. This research also made use of
data and/or software provided by the High Energy Astrophysics Science
Archive Research Center (HEASARC), which is a service of the
Astrophysics Science Division at NASA/GSFC and the High Energy
Astrophysics Division of the Smithsonian Astrophysical Observatory. DA
acknowledges support from the Royal Society.  SG acknowledges the
support of the Centre National d'Etudes Spatiales (CNES).  We thank
the anonymous referee for their efforts.

\facility{NICER, ADS, HEASARC}


\bibliographystyle{aasjournal}

\bibliography{ms}

\newpage


\begin{figure*}
\begin{center}
\includegraphics[scale=0.75]{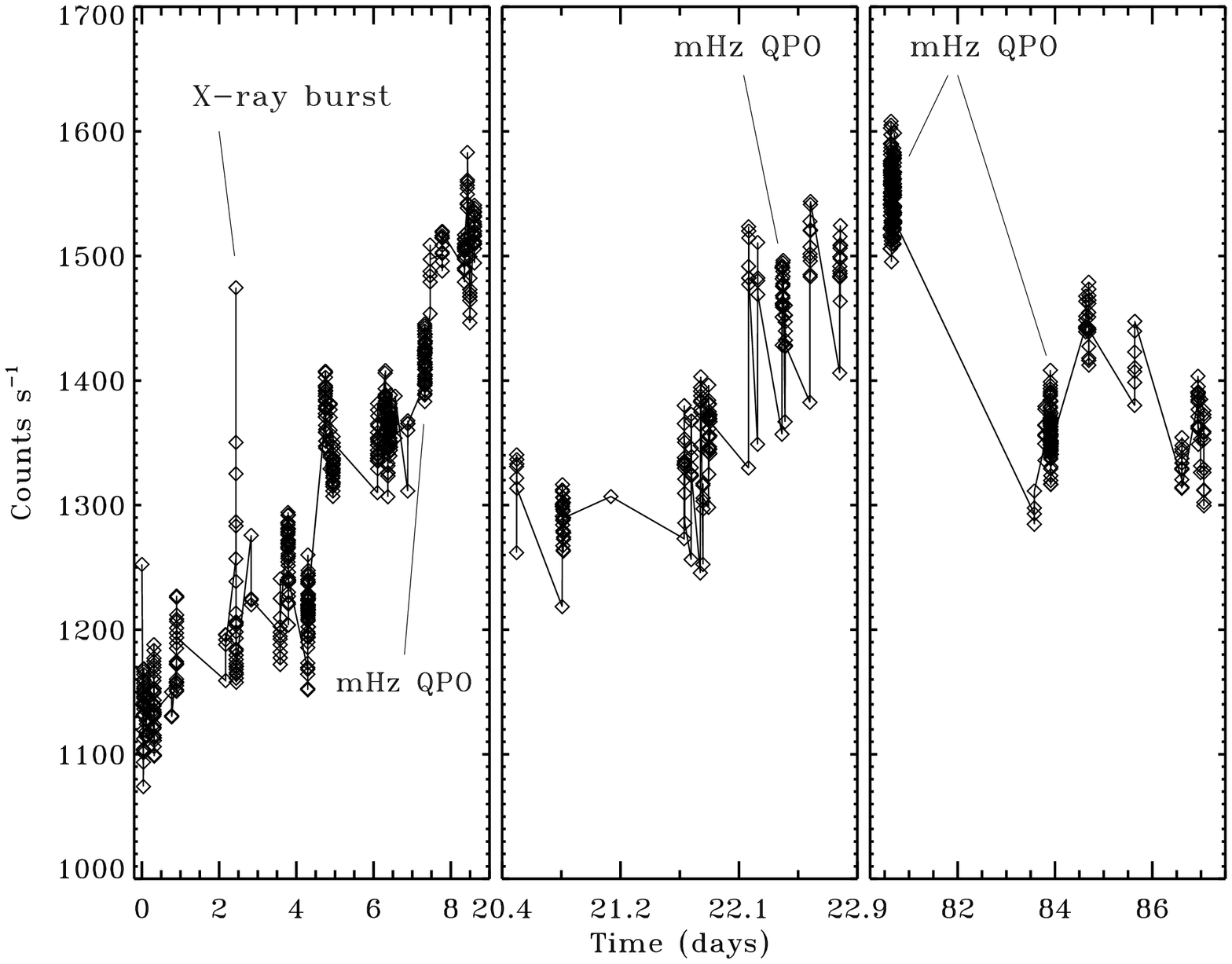}%
\end{center}
\caption{\label{fig:lc} Light curve of \source{} from {\it NICER}
  observations obtained in 2017 June, July and September.  Count rates
  are computed in 32 s bins including events in the 0.4 - 7.5 keV
  band. The time axis is broken up into three panels, one each for the
  June, July, and September exposures, respectively. Time zero
  corresponds to MJD 57924.54154 (TT). Intervals containing an X-ray
  burst and mHz oscillations are indicated. }
\end{figure*}


\begin{figure*}
\begin{center}
\includegraphics[scale=0.75]{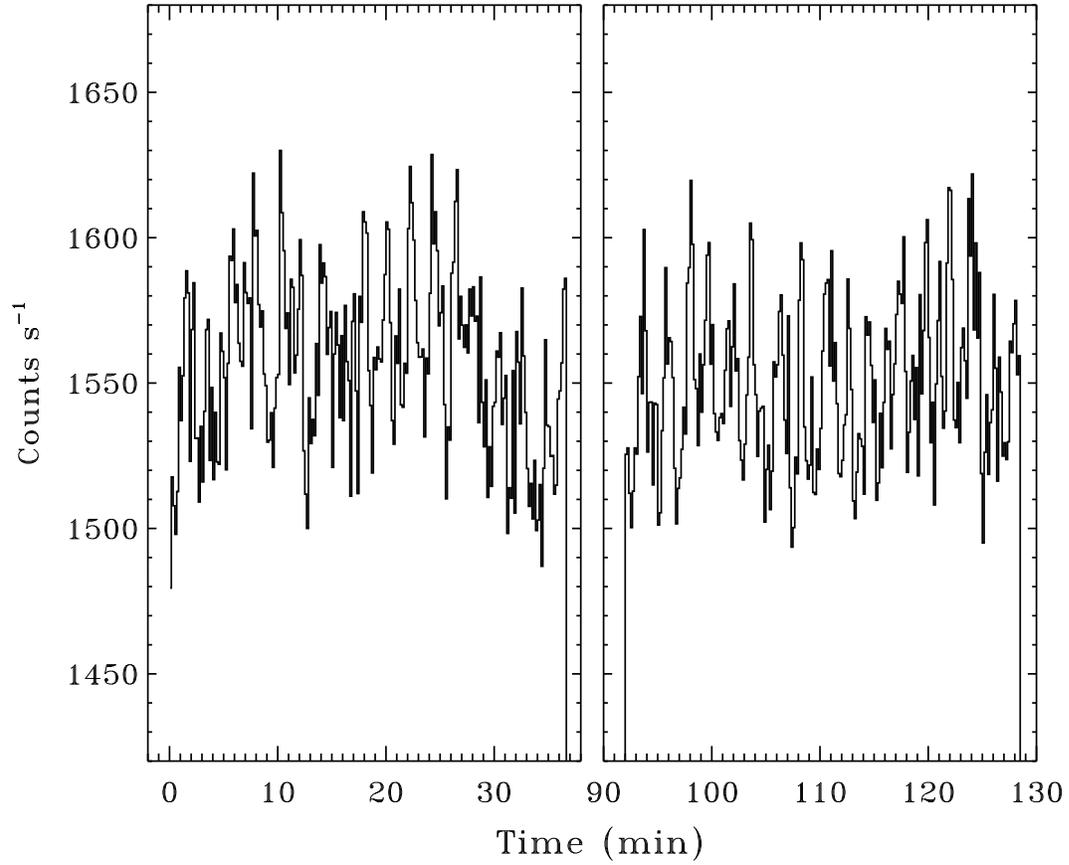}%
\end{center}
\caption{\label{fig:lc} Light curves of \source{} from {\it
    NICER} observations over two consecutive {\it ISS} orbits on 2017
  September 9. The count rates are computed in 10 s bins including
  events in the 0.4 - 7.5 keV band. Pulses with a period near 2 min
  can clearly be seen.  Time zero corresponds to MJD 58005.11848988
  (TT).}

\end{figure*}


\begin{figure*}
\begin{center}
\includegraphics[scale=0.8]{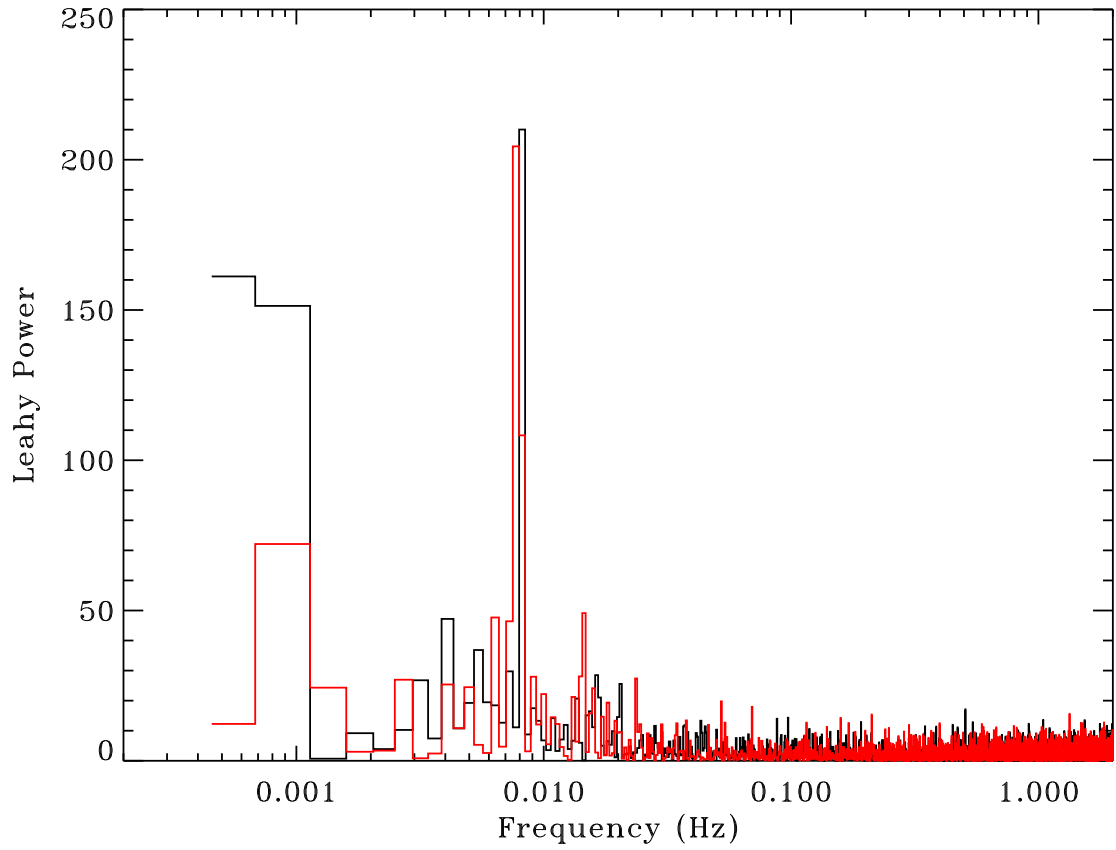}%
\end{center}
\caption{\label{fig:pds_dwells} Power spectra of \source{} from
  the two {\it NICER} dwells shown in Figure 2. The black and red
  traces correspond to the spectra from the first and second dwells,
  respectively.  Strong, narrow peaks are evident near 8 mHz from both
  dwells. }

\end{figure*}


\begin{figure*}
\begin{center}
\includegraphics[scale=0.75]{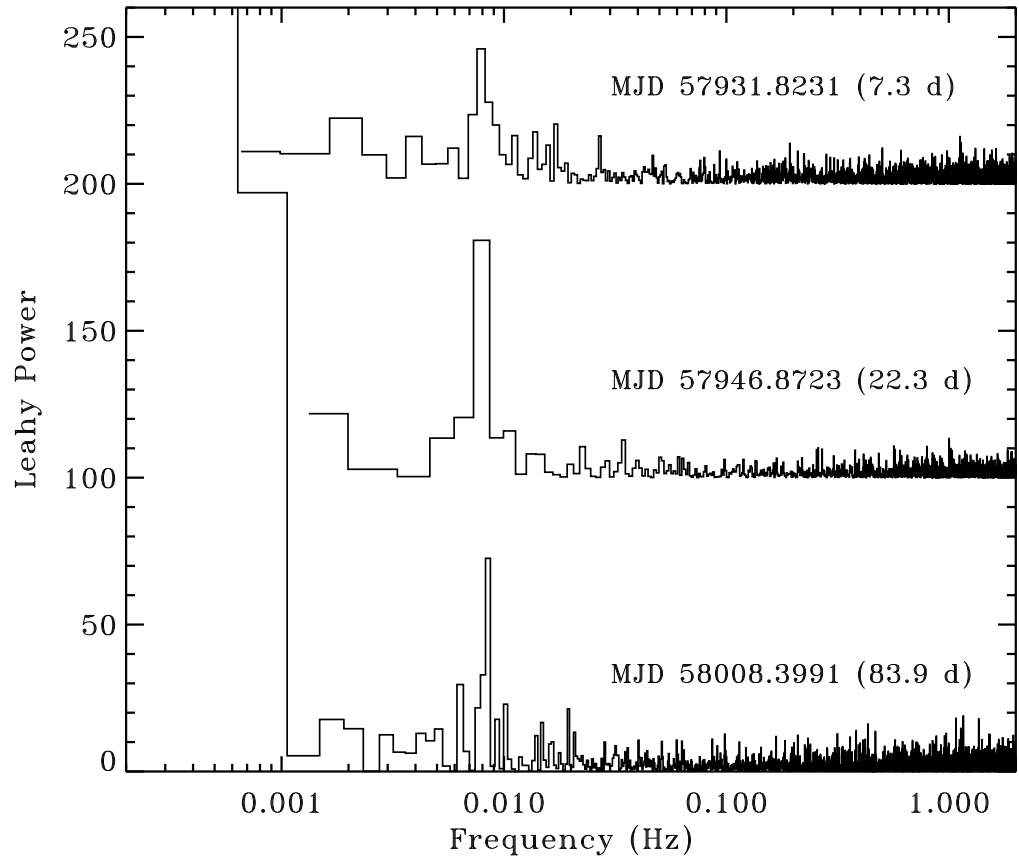}%
\end{center}
\caption{\label{fig:pds2} Power spectra of \source{} from three
  additional {\it NICER} dwells with mHz oscillations. As in Figure 2,
  significant peaks near 8 mHz are evident in each spectrum. Each
  spectrum is labelled with the start time (MJD) of the dwell, with
  the number in parenthesis indicating the corresponding time (day) in
  Figure 1. Each successive spectrum is displaced vertically by 100
  for clarity. }
\end{figure*}

\begin{figure*}
\begin{center}
\includegraphics[scale=0.75]{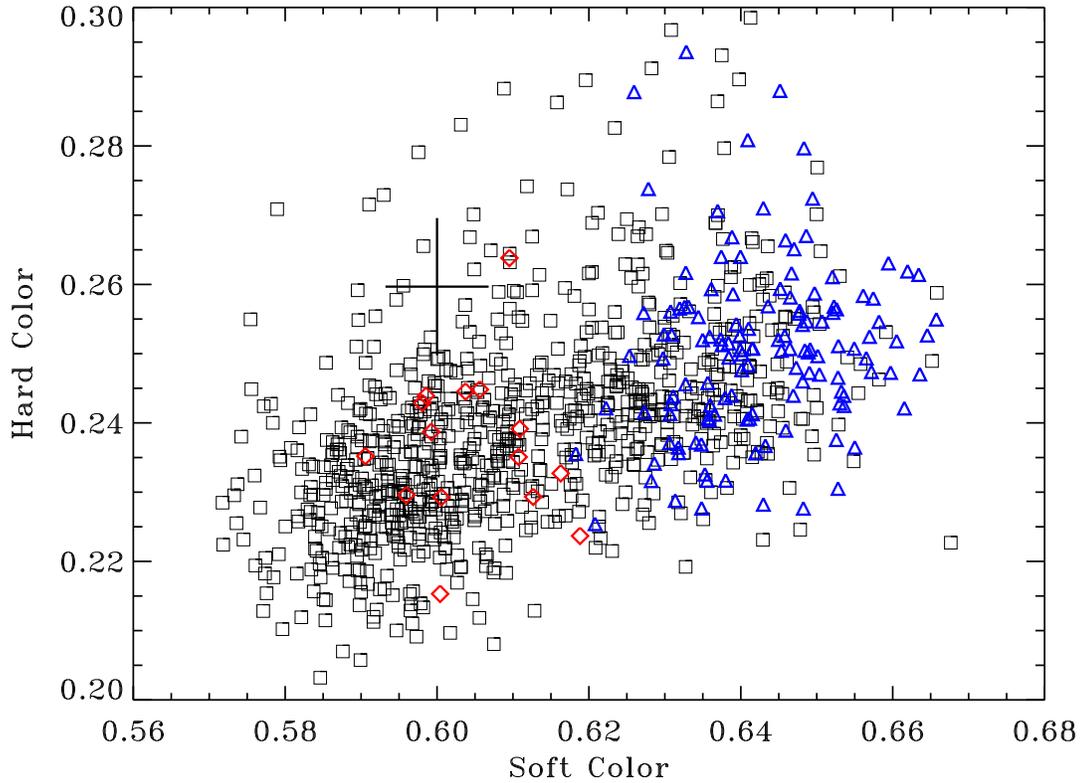}%
\end{center}
\caption{\label{fig:cd} Color - color diagram of \source{} from 34 ks
  of {\it NICER} data obtained in the 2017 June, July and September
  epochs. The hard color is computed as the ratio of count rates in
  the $5.2 - 6.8$ and $3.5 - 5.2$ keV bands.  For the soft color we
  used the $1.8 - 3.5$ and $0.5 - 1.8$ keV bands. Colors were computed
  using 32 s bins. Blue (triangle) symbols were computed from the two
  September epoch dwells (Figure 2) showing mHz oscillations, while
  red (diamond) symbols were computed from intervals immediately
  following the X-ray burst. A typical error bar is also shown.  }
\end{figure*}

\begin{figure*}
\begin{center}
\includegraphics[scale=0.75]{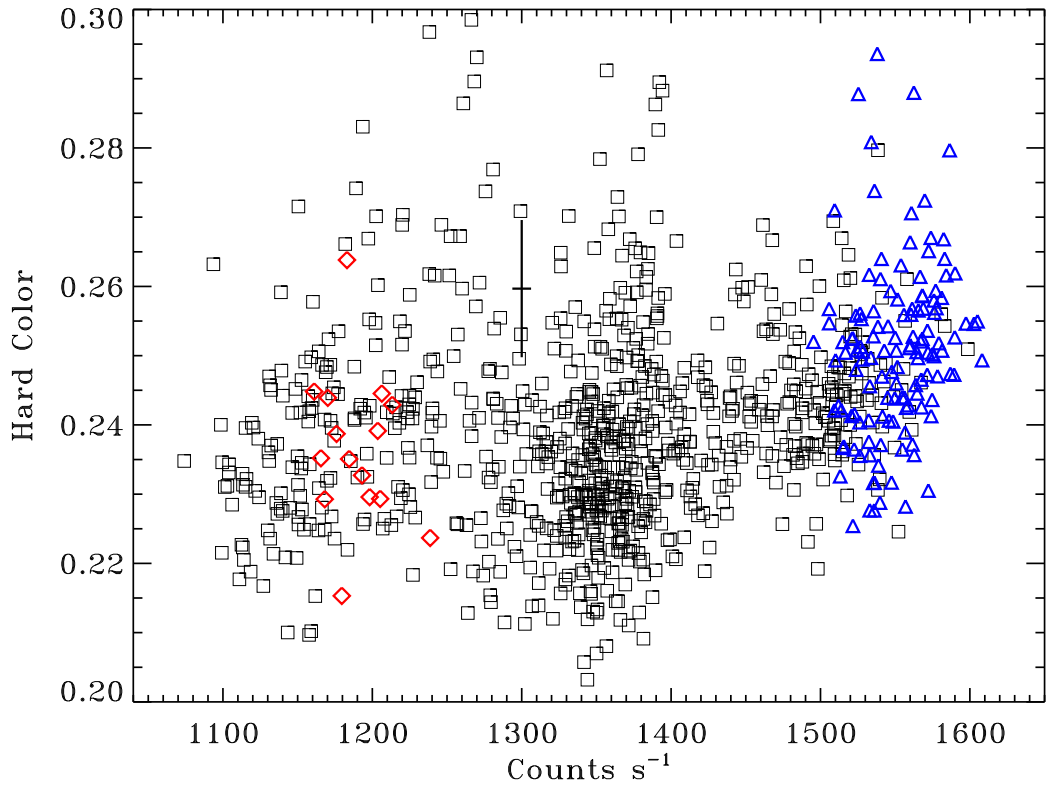}%
\end{center}
\caption{\label{fig:hi} Hardness - intensity diagram of \source{} from
  34 ks of {\it NICER} data obtained in the 2017 June, July and
  September epochs. The hard color is computed as the ratio of count
  rates in the $5.2 - 6.8$ and $3.5 - 5.2$ keV bands.  The intensity
  is computed using the band $0.5 - 6.8$ keV. Values were computed
  using 32 s bins. The colored symbols have the same meanings as in
  Figure 5. A typical error bar is also shown.  }
\end{figure*}

\begin{figure*}
\begin{center}
\includegraphics[scale=0.75]{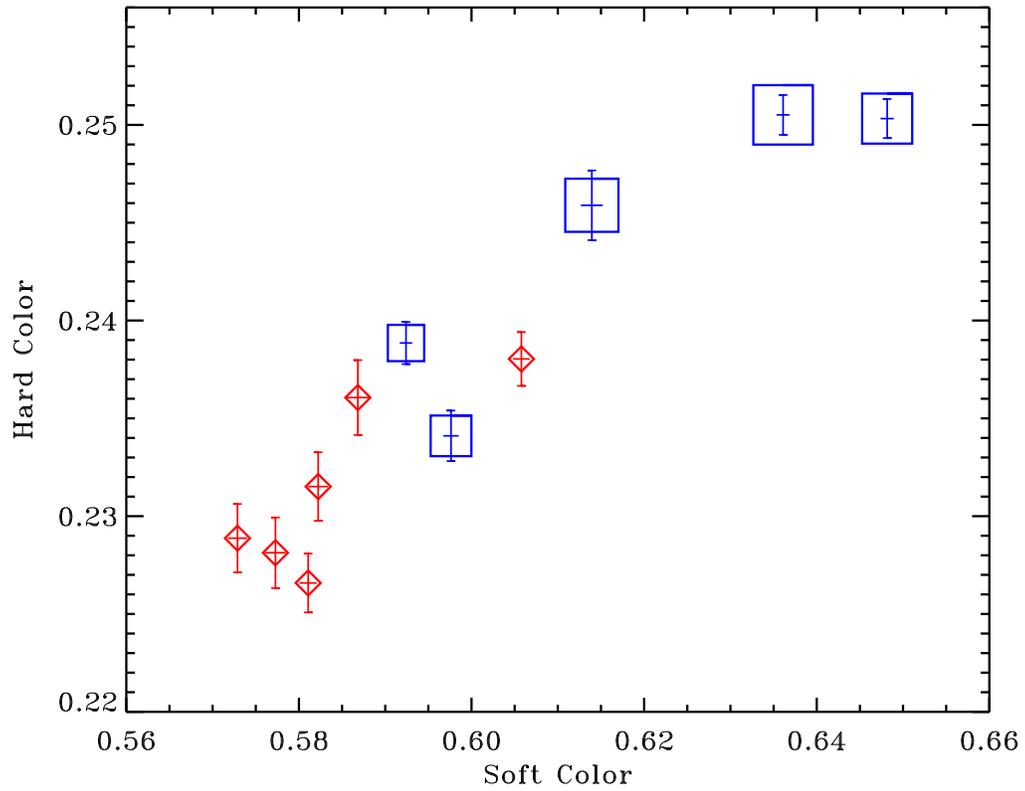}%
\end{center}
\caption{\label{fig:cc_amp} Color - color diagram of \source{} from
  {\it NICER} dwells in which mHz oscillations were detected (blue
  square symbols), as well as several dwells for which no mHz
  oscillations were found (red diamond symbols). For the red diamond
  symbols, the upper limits to the fractional amplitude (rms) are in
  the range from $0.4$ - $0.5 \%$.  For the dwells with mHz
  detections, the symbol size is proportional to the fractional
  amplitude (rms).   }
\end{figure*}

\begin{figure*}
\begin{center}
\includegraphics[scale=0.75]{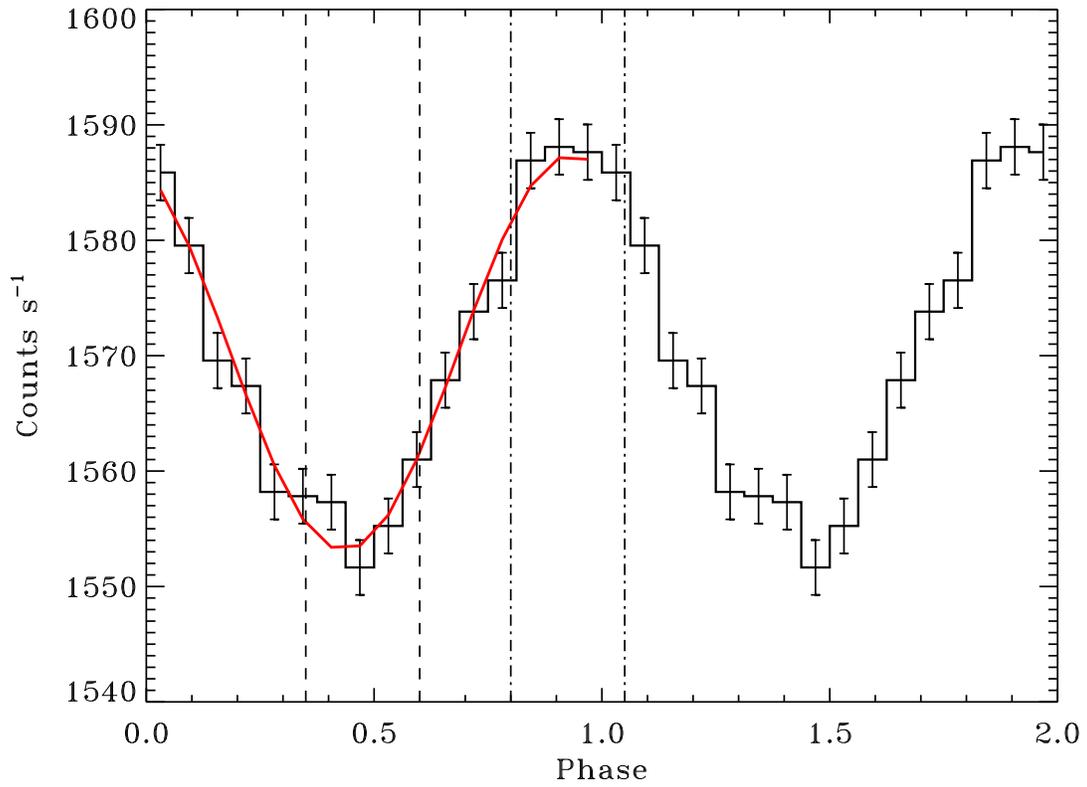}%
\end{center}
\caption{\label{fig:phase_delays} Average, phase-folded mHz
  oscillation profile from the two {\it NICER} dwells shown in Figure
  2. The profile includes all events in the $0.4 - 7.5$ keV band. The
  best fitting model, $A + B\sin(\phi -\phi_0)$, is also shown (red
  curve).  Phase ranges used to extract spectra are indicated by the
  vertical lines. See \S 3 for additional discussion. }
\end{figure*}

\begin{figure*}
\begin{center}
\includegraphics[scale=0.75]{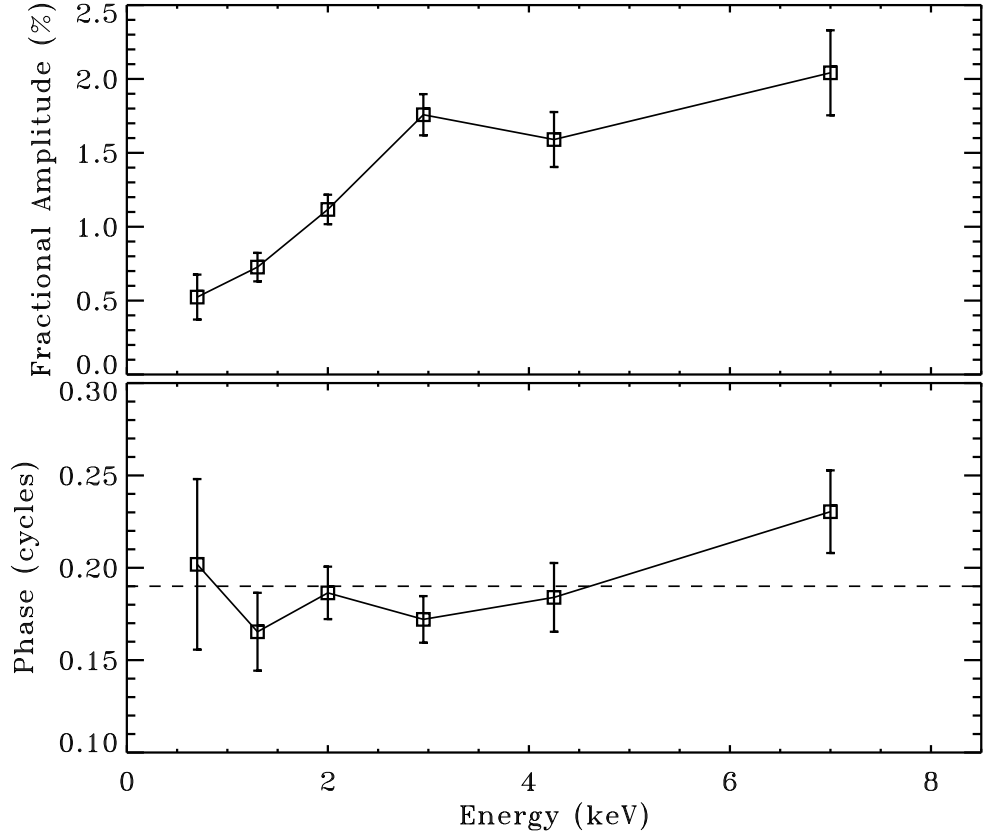}%
\end{center}
\caption{\label{fig:pamps} The fractional pulsed amplitude (top) and
  reference phase, $\phi_0$ (bottom), obtained from fits to the
  phase-folded mHz oscillation profile are shown as a function of
  energy. The dashed line in the bottom panel marks the mean reference
  phase value. This constant value provides an acceptable fit to the
  distribution of values, with a $\chi^2 = 6.9$ for 5 degrees of
  freedom. }
\end{figure*}

\begin{figure*}
\begin{center}
\includegraphics[scale=0.75]{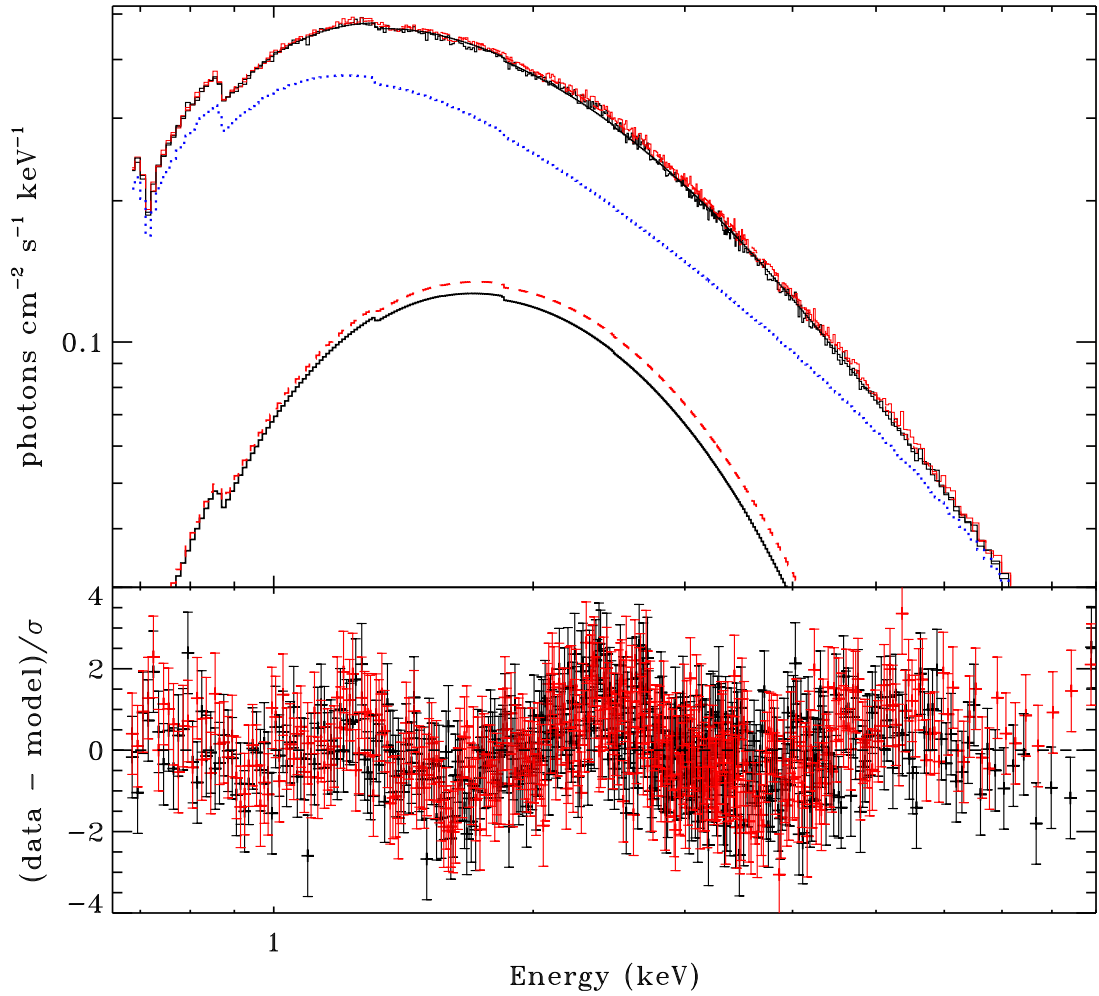}%
\end{center}
\caption{\label{fig:specfits} Phase-resolved {\it NICER} spectra
  extracted from the minimum and maximum mHz oscillation phases
  indicated in Figure 6. The unfolded photon spectra are shown along
  with the total and individual model components (top). The dotted
  curve between the upper and lower curves is the {\it comptt} model
  component, while the two lower curves are the black body {\it
    bbodyrad} components. The red and black curves refer to the
  maximum and minimum phase intervals, respectively. The fit
  residuals, in units of (data $-$ model)/ $\sigma$ are also shown
  (bottom). See discussion in \S 3 for further details. }
\end{figure*}

%
%
%

\end{document}